\documentclass[aps,prb,reprint,showpacs]{revtex4-2}
\usepackage{graphicx} 
\usepackage{amsmath} 
\usepackage{amssymb}
\usepackage{mathtools} 
\usepackage{textcomp}
\usepackage{bm} 
\usepackage{natbib}
\usepackage{natmove}
\usepackage[colorlinks=true, linkcolor=blue, citecolor=blue, urlcolor=blue, linktoc=page, bookmarks=false, pdfstartview={FitH}, pdfborder={0 0 0.0 [3 3]}]{hyperref} 
\usepackage{color} 
\usepackage{dcolumn} 
\newcolumntype{.}{D{.}{.}{2.1}}
\newcolumntype{-}{D{.}{.}{4.0}}
\usepackage{makecell}
\usepackage{multirow}
\usepackage{cleveref}
\usepackage{easyReview} 
\crefname{figure}{Fig.}{Figs}
\Crefname{figure}{Figure}{Figures}
\crefname{table}{Table}{Tables}
\crefname{equation}{Eq.}{Eqs.}
\crefname{section}{Sec.}{Secs.}
\renewcommand{\today}{\number\day \space \ifcase \month \or January\or February\or March\or April\or May\or June\or July\or August\or September\or October\or November\or December\fi \space \number\year} 
\begin{document}
\title{Strain as a topological selector in altermagnetic CrSb}
\author{Sumohan \surname{Giri}}
\email[Contact author: ]{sumohan23@iiserb.ac.in}
\author{Nirmal \surname{Ganguli}}
\email[Contact author: ]{NGanguli@iiserb.ac.in}
\affiliation{Department of Physics, \href{https://ror.org/02rb21j89}{Indian Institute of Science Education and Research Bhopal}, Bhauri, Bhopal 462066, India}
\date{\today}
\begin{abstract}
Altermagnetism combines fully compensated magnetic order with a magnetic symmetry that relates inequivalent spin sublattices, offering a promising, still underexplored platform for unconventional topological phases. Here we show that both isotropic tensile strain and electron localization, controlled by an effective Hubbard interaction $U_{\text{eff}}$, can act as efficient and systematic topological control parameters in the altermagnetic Weyl semimetal CrSb. While CrSb hosts Weyl fermions at equilibrium, modest tensile strain of 4--5\% stabilizes additional symmetry allowed Dirac crossings and triple-point fermions, with further strain selectively favoring the triple-point phase. We propose a 3D low-energy Hamiltonian that captures the interplay between the Hubbard interaction $U$ and the sublattice symmetry of the altermagnet, giving rise to an interaction-driven Dirac crossing. Our results establish CrSb as a model altermagnet in which either strain or electron localization can selectively access and control the distinct topologies inherent to the altermagnets.
\end{abstract}
\maketitle

\section{\label{sec:intro}Introduction}
In altermagnetic materials, the magnetic sublattices compensate exactly, yielding zero net magnetization, yet the electronic bands exhibit strong momentum-dependent spin splitting driven purely by exchange interactions. This unconventional symmetry landscape allows altermagnets to host sizable spin polarization even in the absence of spin-orbit coupling (SOC), setting them apart from both collinear antiferromagnets and traditional magnetic metals. The unusual symmetry breaking present in altermagnets has sparked growing interest in their potential to realize topological electronic phases. Several theoretical works have suggested that the combination of compensated magnetism and nonsymmorphic or crystalline symmetries may naturally stabilize nontrivial quasiparticles---including Dirac, Weyl, and other symmetry-enforced or symmetry-allowed band degeneracies. In particular, Weyl semimetals arising in magnetic systems \cite{Kuroda2017,Liu2019,PhysRevLett.123.187201,Belopolski2019} continue to draw attention due to their robust surface Fermi arcs, chiral anomaly induced transport signatures, and a variety of other unconventional responses associated with Berry curvature.

Although altermagnets have attracted growing interest, the topological aspects of this magnetic class remain comparatively underexplored. CrSb is one notable exception: it has already been identified as an altermagnetic Weyl semimetal \cite{Li2025,Lu2025}. Several studies have examined strain tuning, chemical substitution, and N\'eel-vector reorientation to this compound \cite{PhysRevB.102.224426,PhysRevB.72.064409,Miao2005}. Also magnetization has long been recognized as an effective mechanism for inducing topological phase transitions in a wide range of materials \cite{Du2022,PhysRevX.11.011039}. However, the role of isotropic strain as a systematic topological control parameter in altermagnetic systems has not yet been fully investigated. Motivated by the well-established Weyl semimetal phase in CrSb, our work demonstrates that strain engineering provides a powerful new route to reveal additional topological quasiparticles in this material. By applying modest isotropic tensile strain in the range of $1\% \leq \varepsilon \leq 4\%$, we uncover two additional types of band degeneracies---triple-point (TP) fermions\cite{TP_first,TP_first_Winkler,TP_Chen_Fang,TP_PRL_Winkler,TP_symmorphic} and symmetry-allowed Dirac crossing. These findings establish strain as an effective topological selector in altermagnetic CrSb. Beside Weyl and Dirac topology, these TP fermions have also been shown to give rise to remarkable electronic transport behavior, including ultralow residual resistivity and micron-scale mean free paths in materials such as MoP~\cite{Kumar2019}. Therefore, having triple-point fermions as an additional topological feature in altermagnetic materials represents an important advancement, since such states have not been reported previously in this class of magnetic systems.We then construct a minimal three-dimensional low-energy Hamiltonian for altermagnetic CrSb that captures the combined effects of sublattice symmetry breaking,itinerant antiferromagnetism, and weak spin--orbit coupling.The resulting Bloch Hamiltonian reveals a competition between a symmetry-allowed,momentum-dependent sublattice mass and an interaction-generated exchange field,which drives an insulator--Dirac--insulator sequence upon tuning the effective
Hubbard interaction. This framework provides a unified description of the interaction-controlled Dirac physics observed in altermagnetic materials.Our results therefore position CrSb as a model platform where altermagnetism and strain jointly enable previously unrecognized topological phases.

\begin{figure}
    \centering
    \includegraphics[scale = 0.23]{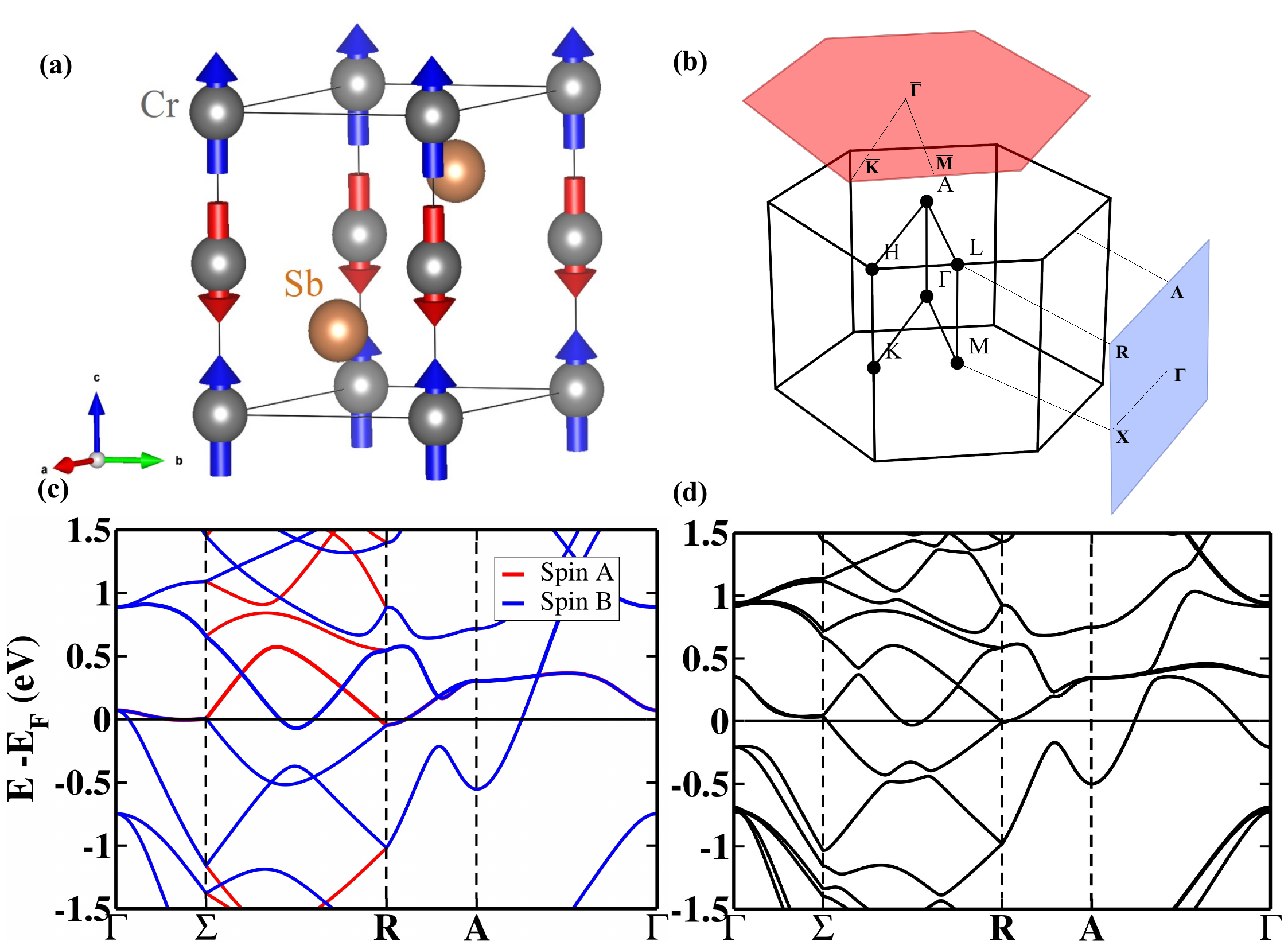}
    \caption{Electronic structure of bulk CrSb with and without spin orbit coupling. (a) The magnetic unit cell of CrSb. (b) Bulk BZ of CrSb with the high symmetry points denoted along with the (001) [red] and (010) [blue] surfaces of the BZ. (c) and (d) Without and with SOC electronic structures of bulk CrSb respectively.}
    \label{fig:unitcell_crsb}
\end{figure}

\section{\label{sec:Method}Methodology}
Our calculations, based on density functional theory (DFT), were performed using the {\scshape vasp} code \cite{vasp, vasp2} within the projector augmented wave (PAW) framework \cite{paw}, combined with a plane-wave basis set having an energy cutoff of 500~eV. The exchange-correlation potential was treated using the local density approximation (LDA) \cite{ldaCA, PerdewPRB81}, and the on-site Coulomb interactions were adjusted through the LDA+$U$ formalism proposed by \citet{DudarevPRB98}. For the unstrained crystal, the value of $U_{\text{eff}}$ was set to 1 eV such that the calculated Cr on-site magnetic moment reproduces some of the experimentally\cite{Cr_moment_exp} and theoretically\cite{Cr_moment_Theory_1,Cr_moment_Theory_2} reported Cr moment in CrSb, providing a reliable starting point for the study of strain-driven topological features. The integration over the Brillouin zone was performed using a $15 \times 15 \times 13$ $\Gamma$-centered $k$-point mesh within the corrected tetrahedron method \cite{BlochlPRB94T}. Spin-orbit interaction (SOI) was included self-consistently in all calculations pertaining to the topological analysis. Isotropic tensile strain was applied by uniformly scaling the in-plane and the out-of-plane lattice parameters while allowing the atomic positions to fully relax until the Hellmann-Feynman force on each atom was below a threshold of $10^{-2}$~eV/\AA.To investigate the topological properties, maximally localized Wannier functions (MLWFs) were constructed for the Cr $3d$, $4s$ and Sb $5p$ orbitals using the {\scshape Wannier90} code \cite{wannier90}. The resulting Wannier interpolated Hamiltonian accurately reproduces the DFT bands near the Fermi level and serves as the basis for calculations of the Berry phase and all surface states. The surface spectral functions were computed using the iterative Green’s function method implemented in the {\scshape WannierTools} package \cite{wanniertool}.
\section{\label{sec:Result}Results and Discussion}
The results of our DFT calculations, symmetry analysis and the description of Hamiltonian are discussed in detail below.
\subsection{Triple-point formation along the A-$\Gamma$ direction}
\begin{figure}
    \centering
    \includegraphics[scale=0.21]{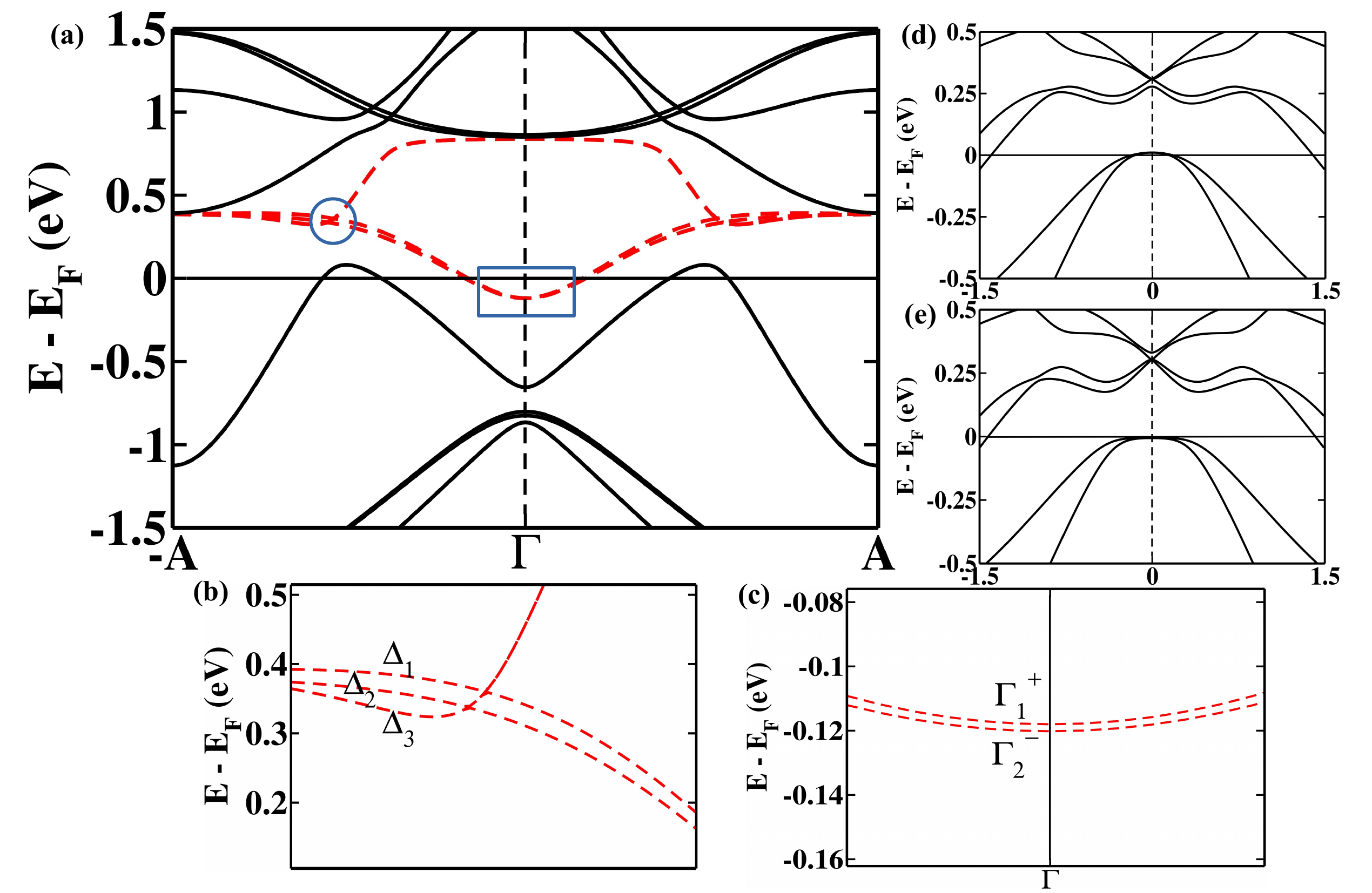}
    \caption{\label{fig:TP}Electronic band structure showing the Triple points and the compatibility relations of the bands. (a) The pair of TPs along the A--$\Gamma$ (A--$\Gamma$) line produced by the four bands labeled as red dashed lines. (b) and (c) Zoomed in portion shown in blue circle and rectangluar box respectively in (a). (d) and (e) Fine tuned pair of TPs at $k_z = -0.271$ and $k_z = -0.281$ inside the BZ along the -A--$\Gamma$ line respectively. The x-axis denotes the $k_x$ line.}
\end{figure}

We first investigate the electronic structure of hexagonal CrSb using \textit{ab initio} calculations.The compound crystallizes in the NiAs-type structure\cite{SnowPR52}shown in \cref{fig:unitcell_crsb}(a) with space group $P6_{3}/mmc$ (No.~194). The primitive unit cell contains two atoms, Cr and Sb, occupying the Wyckoff positions $2a$ $(0, 0, 0)$ and $2c$ $(\frac{1}{3}, \frac{2}{3}, \frac{1}{4})$ respectively. The corresponding bulk Brillouin zone (BZ) and its high-symmetry paths are shown in \cref{fig:unitcell_crsb}(b). In the absence of spin--orbit coupling(SOC), the band structure exhibits a characteristic sixfold degenerate crossing along the $C_{3v}$-symmetric A--$\Gamma$ line \cref{fig:unitcell_crsb}(c). The presence of the continuous spin-rotation symmetry $U_z(\phi)$, with $\phi \in [0,2\pi)$, in the spin-only group of collinear CrSb \cite{unconventional_magnons}, together with the anti-unitary symmetry $TA$ or the combined symmetry $U_n(\pi)A$ (where $A$ represents a pure lattice-space operation) in the spin space little co-group $^{2}6/^{m}m^{1}m^{2}m^{\infty}1$ \cite{Qihang_Liu_SSG}, ensures that all bands become twofold degenerate along the A--$\Gamma$ high symmetry line.
Furthermore, a fourfold-degenerate band intersects with such a twofold-degenerate band, giving rise to a sixfold-degenerate quasiparticle along the A--$\Gamma$ direction.Upon inclusion of SOC, the magnetic little group along the A--$\Gamma$ line becomes $-6'm2'$ \cite{Barnevig_TQC_1,Barnevig_TQC_2}. The corresponding co-little group contains the symmetry elements of the $C_{3v}$ point group, namely the threefold rotation $C_{3}$ and the three vertical mirror planes $\sigma_{v}$, each containing the $C_{3}$ rotation axis and rotated by $120^{\circ}$ relative to one another.As discussed by Winkler \textit{et al.} \cite{Winkler2019}, such a symmetry setting can in principle host triple-point fermions. However, in unstrained CrSb we do not observe any clear triple-point fermions along the A--$\Gamma$ direction Fig.~\ref{fig:unitcell_crsb}(d). Upon applying isotropic tensile strain and gradually increasing it from $1\%$ to about $4\%$, we find that a distinct triple-point-like crossing emerges, as illustrated in \cref{fig:TP}(a). Increasing the on-site interaction strength $U_{\text{eff}}$ from 1 eV to 3.5 eV produces an analogous effect, similarly stabilizing the triple-point fermions along the A--$\Gamma$ line. Figures \cref{fig:TP}(d) and \cref{fig:TP}(e) shows the dispersions in the (100) direction for kz tuned to the positions of pair of TPs shown in the \cref{fig:TP}(a). Moreover, upon further increasing either $U_{\mathrm{eff}}$ or the tensile strain, these triple-point crossings remain robust and continue to persist along the same high-symmetry direction. This persistence indicates that the TPs are protected as long as the applied perturbations preserve the underlying crystalline and magnetic symmetries of the little group. A summary of the evolution of this triple-point features under these symmetry-preserving tuning parameters is given in \cref{tab:UandStrain}.To elucidate the symmetry origin of the pair of triple points appearing along the A--$\Gamma$ direction, we examine the evolution of the relevant band representations under spin--orbit coupling (SOC).Upon including SOC, the band representations are most conveniently described starting from the A point \cite{Barnevig_TQC_1,Barnevig_TQC_2}. The twofold-degenerate representations $A_{1}$ and $A_{3}$ evolve along the A--$\Gamma$ line into the $\Delta$ branches.Specifically, $A_{3}$ splits into the singly degenerate $\Delta_{1}$ and $\Delta_{2}$ bands, while $A_{1}$ evolves into the doubly degenerate $\Delta_{3}$ branch as seen in~\cref{fig:TP}(b).Along the A--$\Gamma$ direction, the $\Delta_{3}$ branch intersects both $\Delta_{1}$ and $\Delta_{2}$, and these symmetry-allowed band crossings give rise to a pair of triply degenerate points.Near the $\Gamma$ point, the $\Delta_{1,2}$ states connect to $\Gamma_{1}^{+}$ and $\Gamma_{2}^{-}$[see~\cref{fig:TP}(c)], while $\Delta_{3}$ connects to $\Gamma_{3}^{+}$. This connectivity reflects the SOC-induced lifting of the original sixfold degeneracy and the emergence of type-B triple point fermions in strained CrSb.

\subsection{Dirac-like band dispersion and its tunability via strain and electron localization in CrSb}
In the absence of spin--orbit coupling (SOC), the band structure of CrSb exhibits a distinct fourfold degenerate Dirac-like crossing along the L--A direction, as shown in \cref{fig:unitcell_crsb}(c). The linear dispersion around this crossing indicates the presence of massless Dirac fermions. This feature is consistent with the spin space group symmetry analysis \cite{Qihang_Liu_SSG,SSG_irreps_application}, that the extra spin degeneracy and the antiunitary symmetries it contains,allows the four fold degeneracy to occur along high-symmetry lines in the hexagonal Brillouin zone. Upon inclusion of SOC, this Dirac point becomes gapped due to the lifting of the spin degeneracy, transforming the system into a trivial semimetal.

\begin{figure*}
    \centering
    \includegraphics[scale=0.32]{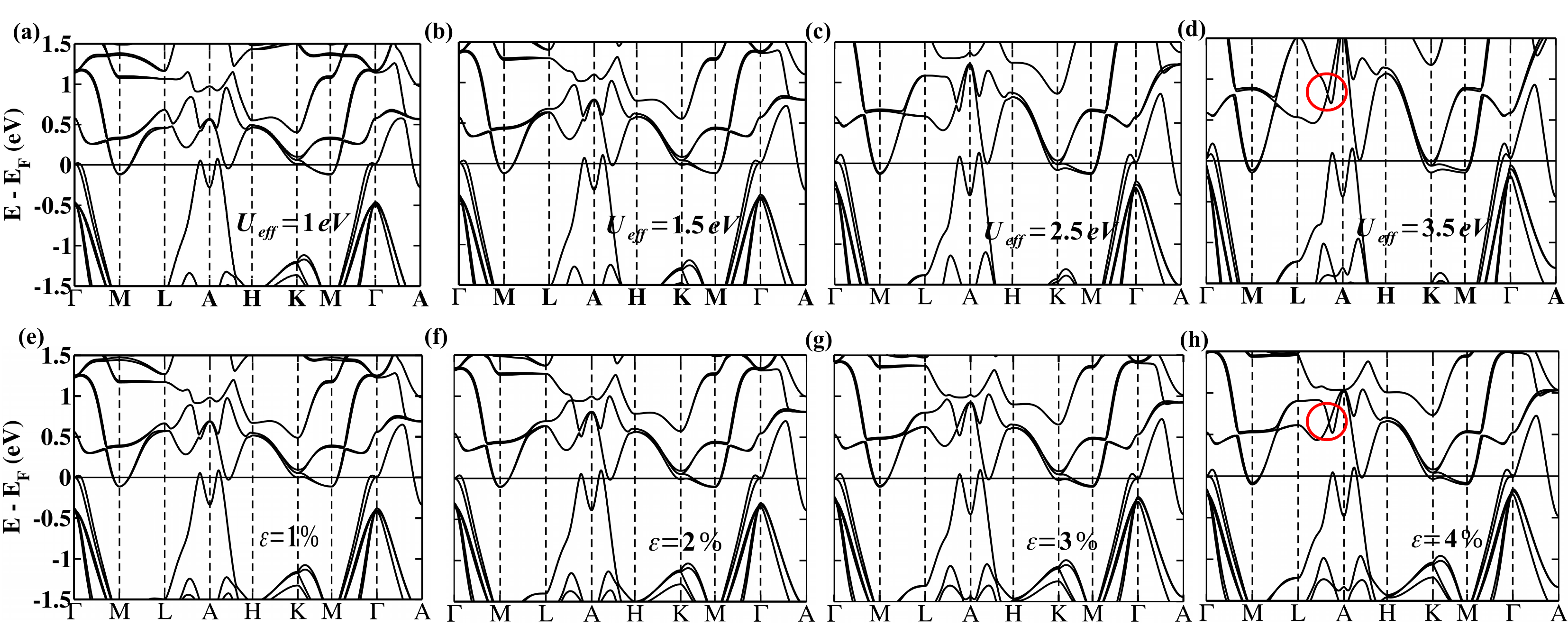}
    \caption{Electronic band structures calculated with different $U_{\text{eff}}$ values and also with isotropic tensile strain. (a)--(d) The electronic band structures of CrSb calculated with $U_{\text{eff}} =$ 1 eV, 1.5 eV, 2.5 eV, and 3.5 eV respectively. It shows the evolution of the Dirac crossing from $U_{\text{eff}} =$ 1 eV to nearly 3.5 eV. (e)--(h) The calculated electronic band structure with isotropic tensile strain starting from 1\%, and then 2\%, 3\%, and 4\% respectively. In all of the above strained case we took the $U_{\text{eff}}$ value to be 1 eV.  It shows the same evolution of the Dirac crossing as in the varying $U_{\text{eff}}$ case. The red circles on (d) and (h) shows the observed Dirac crossing at 3.5 eV $U_{\text{eff}}$ value and 4\% isotropic tensile strain respectively.}
    \label{fig:Dirac}
\end{figure*}

To further probe the stability and tunability of this Dirac-like state, we employed strain engineering. Remarkably, when a isotropic tensile strain is applied, the Dirac-like crossing re-emerges for strain values reaching to 4\%  \cref{fig:Dirac}(e)-\cref{fig:Dirac}(h).A similar effect can be achieved by increasing the on-site Coulomb repulsion $U_{\text{eff}}$ within the Cr $3d$  as seen from our DFT calculations \cref{fig:Dirac}(a)-\cref{fig:Dirac}(d),introduced in the LDA+$U$ framework. Enhancing $U_{\text{eff}}$ increases the localization of the $3d$ electrons, effectively lowering the hopping strength $t$ and at the same time increases the onsite local magnetic moment of Cr \cite{TSD_LDA+U}. Both tensile strain and increasing $U_\text{eff}$ push the system toward a regime where electronic localization competes with band dispersion, consistent with the heuristic picture of a changing $U/t$ ratio commonly invoked in correlated Dirac systems. As summarized in \cref{tab:UandStrain}, the Dirac crossing on the L--A line appears only within a finite window of tensile strain, whereas increasing $U_{\mathrm{eff}}$ continues to preserve the Dirac crossing once it is established. Under moderate tensile strain (approximately $4$--$6\%$), the reorganization of the Cr $3d$ states places irreducible representations of different symmetry in contact along L--A, producing a symmetry-protected Dirac point. However, further increasing the strain alters the orbital hierarchy and hopping amplitudes, separating the bands and opening a gap. This behavior is reminiscent of the evolution reported in artificial graphene and engineered surface lattices, where electron--electron interactions primarily tune a single-band tight-binding model. In such systems, the physics of Dirac cones is governed by a single effective hopping integral $t$, and varying $U/t$ (e.g., via deeper wells, larger dot separations, or increased occupation) directly drives the transition between localized, delocalized, and gapped regimes \cite{PRL-U/t, nature_U/t}. In contrast, CrSb is a multi-orbital material involving several Cr $3d$ and Sb $p$ states, substantial crystal-field splitting, and multiple irreducible representations controlling band connectivity near the Fermi level. Therefore, there is no single scalar $t$ or single orbital degree of freedom that determines the Dirac point. Instead, the emergence of the Dirac crossing in CrSb reflects a more complex balance of (i) electronic localization \cite{correlation-localization}, (ii) bandwidth renormalization \cite{renormalization-localization}, and (iii) strain-driven reordering of orbital energies and hybridization \cite{strain-orbital-effect_1,strain-orbital-effect_2,strain-orbital-effect_3}. 

\begin{figure}
    \centering    
    \includegraphics[scale=0.38]{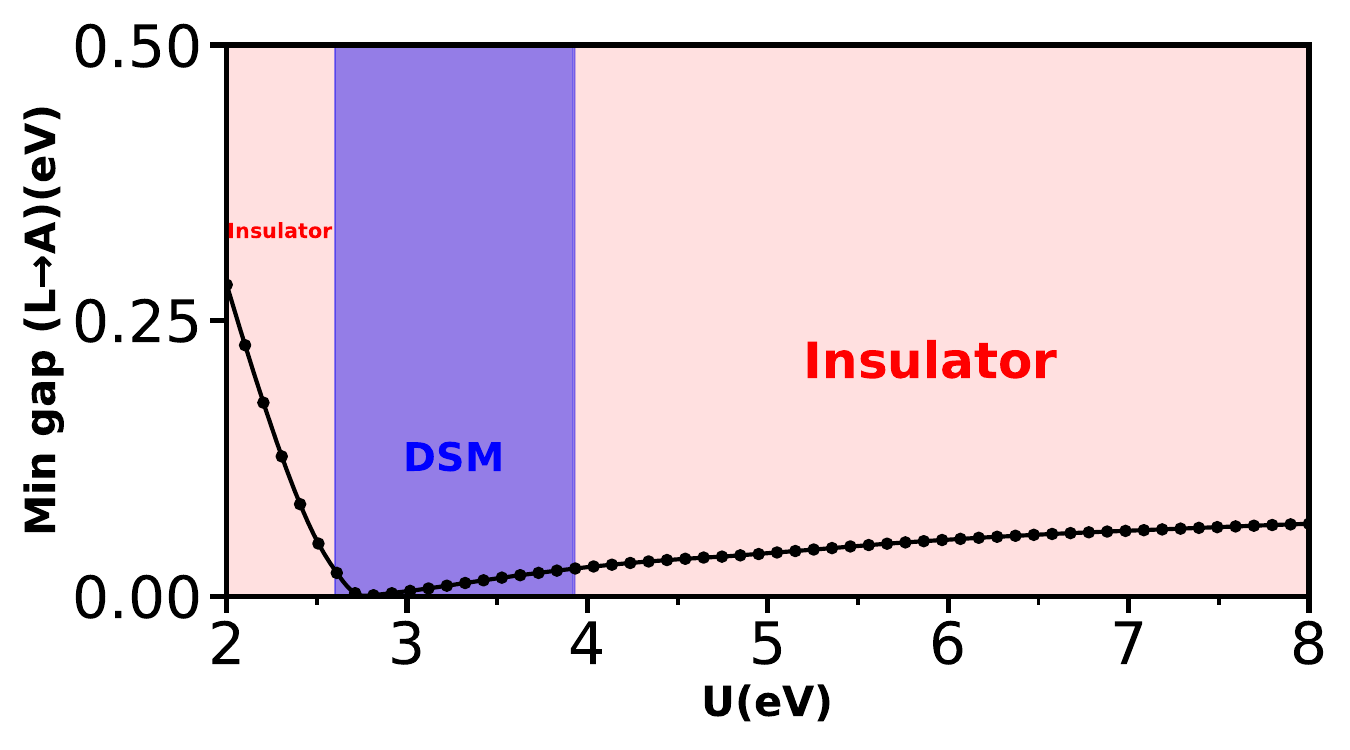}
    \caption{Evolution of the minimum bulk band gap along the L--A line as a function of theeffective interaction strength $U$, obtained from the low-energyBloch Hamiltonian derived in the ~\cref{simplified_H}. For small $U$, the system isinsulating due to a finite, symmetry-allowed sublattice mass. Upon increasing $U$, the interaction-generated exchange field compensates the band mass at a critical value, leading to a gap closing and the emergence of a Dirac semimetal phase. For larger $U$, the gap reopens, yielding a second insulating phase.}
    \label{fig:gap_vs_U}
\end{figure}

To explain this complex interplay between U and band topology we construct a 3D effective Bloch Hamiltonian with the Hubbard hamiltonian along with the nearest neighbour sublattice hoppings
\begin{align}
H(\vec{k}) &=
\epsilon(\vec{k})\, \mathbb{I}_2 \otimes \tau_x
+
\alpha\, M_x(\vec{k})\, \sigma_x \otimes \tau_z
+
\alpha\, M_y(\vec{k})\, \sigma_y \otimes \tau_z
\nonumber\\
&\quad
+
\big(\Delta - M(\vec{k})\big)\, \sigma_z \otimes \tau_z
+
\eta\, \mathbb{I}_4.
\label{simplified_H}
\end{align}
where $\epsilon(\vec{k})$ arises from out-of-plane nearest- and second--nearest-neighbor hopping, $M_{x,y}(\vec{k})$ originate from intrinsic spin--orbit coupling, $M(\vec{k})$ is a momentum-dependent sublattice-odd band mass, and $\Delta$ denotes the interaction-generated antiferromagnetic exchange field.The term $\eta\,\mathbb{I}_4$ represents a uniform energy shift that fixes the Dirac crossing at a finite energy away from the Fermi level and does not affect the band topology or symmetry properties of the Hamiltonian. Crucially, in CrSb the magnetic space-group symmetry is altermagnetic and admits a spin-alteration operation that exchanges the two sublattices. As a result, the sublattice-odd mass $M(\vec{k})$ is symmetry allowed and generically nonzero throughout the Brillouin zone, rendering the noninteracting system insulating. This nonvanishing band mass originates from inequivalent onsite energies and second--nearest-neighbor hopping amplitudes on the two magnetic sublattices, which are not constrained to be equal by symmetry.
Including electron--electron interactions at the itinerant mean-field level generates a staggered antiferromagnetic exchange field $\Delta = \frac{U}{2} m_z$, which enters the Hamiltonian in the same symmetry channel as the band mass $M(\vec{k})$. The effective gap is therefore controlled by the difference $\Delta - M(\vec{k})$. For weak interactions, $\Delta < M_{\min}$ and the system remains fully gapped. Upon increasing the Hubbard interaction strength, $\Delta$ can compensate the pre-existing band mass at symmetry-enforced momenta where $\epsilon(\vec{k}) = 0$. Owing to the negligibly weak spin--orbit coupling in CrSb, the conditions $M_x(\vec{k}) = M_y(\vec{k}) \approx 0$ are satisfied to an excellent approximation, leading to an exact closing of the bulk gap and the formation of a three-dimensional Dirac point. For larger interaction strength, $\Delta > M_{\max}$, the effective mass changes sign and the gap reopens, yielding a second insulating phase. The resulting evolution of the band gap with increasing interaction strength is shown in Fig.~\cref{fig:gap_vs_U}. More generally, if a given high-symmetry point or line in the Brillouin zone exhibits an insulator--Dirac--insulator transition upon tuning $U_{\mathrm{eff}}$, this behavior can be understood directly from the interplay between the interaction-generated exchange field and the momentum-dependent band mass. In the present case, this mechanism operates along the L--A line, for weak interactions the condition $\Delta < M_{\min}$ ensures a finite gap, while at a critical interaction strength the equality $\Delta = M(\vec{k})$ is satisfied at a symmetry-enforced momentum, resulting in a Dirac point. Upon further increasing the interaction strength such that $\Delta > M_{\max}$, the effective mass no longer vanishes anywhere along L--A, and the system re-enters an insulating phase.

\subsection{Group-theoretical analysis of the strain-induced Dirac degeneracy along L--A}

To establish the symmetry origin of the strain-induced Dirac crossing along the L--A line in CrSb, we analyze the magnetic little co-group of this line. In the magnetic configuration considered here, the little co-group belongs to $2^{\prime}2^{\prime}2$, generated by the elements $\{E\,|\,0\}$, $\{2_{001}\,|\,0\}$, $\{2^{\prime}_{100}\,|\,0\}$, and $\{2^{\prime}_{010}\,|\,0\}$. This nonunitary magnetic co-group admits two inequivalent two-dimensional co-representations, denoted as $\text{R1R2}(2)$ and $\overline{\text{R}3\text{R}4}(2)$ in the magnetic representation notation \cite{Barnevig_TQC_1,Barnevig_TQC_2}, appropriate for this high symmetry line.

Each representation remains doubly degenerate throughout the L--A line due to the antiunitary symmetry operations $\{2'_{100}\mid 0\}$ and $\{2'_{010}\mid 0\}$, which enforce a Kramers-like degeneracy even in the absence of conventional time-reversal symmetry. Because the two 2D irreducible representations appearing on the L--A line originate from different sets of irreps at the A point, as required by the global compatibility relations of the little group along L--A in the $P6_3'/m'm'c$ magnetic space group (No.\ 194.268) of CrSb, the corresponding basis states must transform differently under the unitary symmetry operation $\{2_{001}\mid 0\}$ with the eigenvalues $\{+i,-i\}$. So the two doublets span distinct invariant subspaces that cannot be coupled by any symmetry-allowed perturbation, preventing hybridization when they are energetically degenerate.
When the energies of $\text{R1R2}(2)$ and $\overline{\text{R}3\text{R}4}(2)$ become equal under tensile strain, the two doubly-degenerate bands meet at the same ${k}$ point, producing a fourfold degeneracy. This crossing is therefore magnetically symmetry allowed: it originates from the accidental coincidence of two independent 2D co-representations of the magnetic little co-group $2^{\prime}2^{\prime}2$, rather than from symmetry breaking. The strain-induced restoration of this resonance between $\text{R1R2}(2)$ and $\overline{\text{R}3\text{R}4}(2)$ thus yields the Dirac-like fourfold node observed in our calculations, fully consistent with the magnetic group theoretical analysis.

\subsection{Topological surface states of Dirac and TP fermions in strained CrSb}

\begin{figure*}
    \centering
    \includegraphics[width=1.02\textwidth]{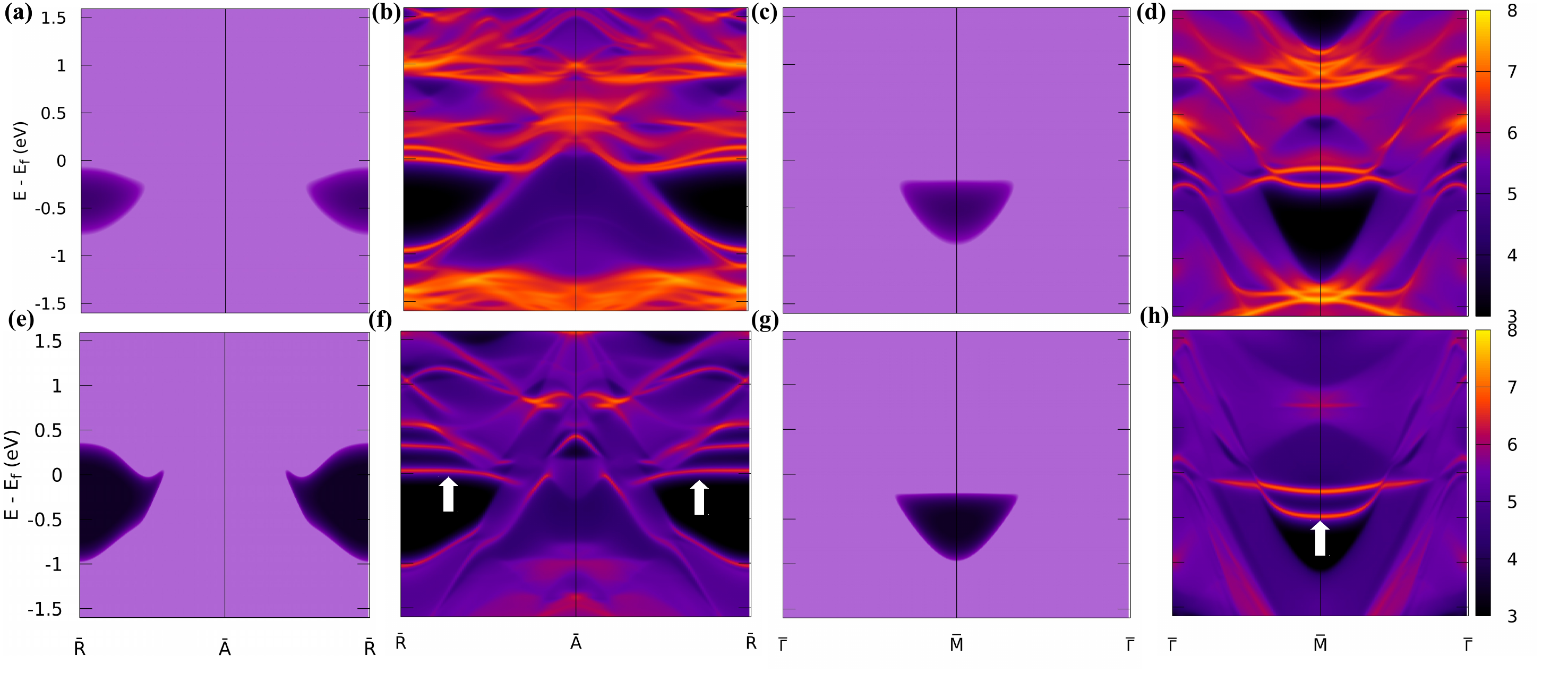}
    \caption{Topological surface state of unstrained and strained CrSb. (a)--(d) The surface states of unstrained CrSb in which (a) and (c) show the bulk projected surface density of states on the (010) and (001) surfaces respectively, while (b) and (d) show the projected surface density of states for the (010) and (001) surface BZ respectively. (e)--(h) The surface states of strained (4\%) CrSb in which (e) and (g) show the bulk projected surface density of states on the (010) and (001) surfaces respectively, and (f) and (h) show the projected surface density of states for the (010) and (001) surfaces of CrSb respectively. In figs. (f) and (h), the vertical white arrows indicate the clear topological surface states that are absent in the unstrained case shown in figs. (b) and (d).}
    \label{fig:surface_states}
\end{figure*}

To establish the topological nature of the strain-stabilized Dirac and TP fermions, we computed the surface spectral functions for the relevant high-symmetry planes of the Brillouin zone. The tensile strain drives the system into a regime where the band crossings along the A–$\Gamma$ and L–A directions acquire nontrivial topology, manifested through the appearance of robust surface states.To further confirm the nontrivial nature of the strain-induced Dirac crossing that simultaneously appears on the L--A line, we evaluated the Berry phase associated with the Dirac point. The resulting Berry phase is approximately $0.8\pi$, indicating a nontrivial topological characteristic of Dirac-like quasiparticles. In contrast, the unstrained structure, which lacks this Dirac crossing, exhibits a trivial Berry phase equal to zero. The continuous evolution of the Berry phase from 0 to $0.8\pi$ under increasing tensile strain thus reveals a clear topological transition accompanying the formation of the Dirac point. To visualize the surface manifestations of these bulk topological features, we computed the surface spectral functions for both the (001) and (010) surfaces for both the unstrained and strained CrSb using the iterative Green's function method applied to the Wannier tight-binding Hamiltonian. In each case, we separately evaluated the bulk-projected band structure and the corresponding surface spectral weight. The bulk projections for the two surface orientations for the unstrained CrSb are shown in \cref{fig:surface_states}(a)-\cref{fig:surface_states}(c), while \cref{fig:surface_states}(b)-\cref{fig:surface_states}(d) display the surface spectra along the $\bar{R}$--$\bar{A}$--$\bar{R}$ and $\bar{\Gamma}$--$\bar{M}$--$\bar{\Gamma}$ directions for the (010) and (001) surfaces, respectively. In both orientations, for the strained case distinct surface bands \cref{fig:surface_states}(f)-\cref{fig:surface_states}(h) emerge inside the bulk energy gaps as seen in the \cref{fig:surface_states}(e)-\cref{fig:surface_states}(g). These clearly visible surface states provide strong evidence that the strain-induced Dirac and TP fermions in CrSb possess a robust topology.

\begin{table}
    \caption{\label{tab:UandStrain}
    Summary of the effect of on-site Coulomb interaction $U_{\text{eff}}$ and tensile strain on the magnetic moment of Cr in CrSb and the emergence of Dirac and triple-point (TP) fermions. Results are obtained from DFT calculations within LSDA+SOC. A black tick ($\checkmark$) indicates the presence of a symmetry-allowed Dirac or TP crossing along the corresponding high-symmetry line, while a black cross ($\times$) denotes its absence.}
\begin{ruledtabular}
\begin{tabular}{l . c c}
Parameter & \multicolumn{1}{c}{Cr moment ($\mu_B$)} & Dirac & TP \\ 
\hline
$U_{\text{eff}} = 1$ eV   & 2.91 & $\times$ & $\times$ \\
$U_{\text{eff}} = 1.5$ eV & 3.01 & $\times$ & $\times$ \\
$U_{\text{eff}} = 2.5$ eV & 3.31 & $\times$ & $\times$ \\
$U_{\text{eff}} = 3.5$ eV & 3.51 & $\checkmark$ &  $\checkmark$ \\
$U_{\text{eff}} = 5.5$ eV & 3.81 & $\checkmark$ & $\checkmark$ \\
$U_{\text{eff}} = 6.5$ eV & 3.93 & $\checkmark$ & $\checkmark$ \\
$U_{\text{eff}} = 8$ eV & 4.08 &$\times$ & $\checkmark$ \\
\hline
Strain = $1\%$  & 3.00 & $\times$ & $\times$ \\
Strain = $2\%$  & 3.07 & $\times$ & $\times$ \\
Strain = $3\%$  & 3.14 & $\times$ & $\times$ \\
Strain = $4\%$  & 3.21 & $\checkmark$ & $\checkmark$ \\
Strain = $6\%$  & 3.33 & $\checkmark$ & $\checkmark$ \\
Strain = $7\%$  & 3.39 & $\times$ & $\checkmark$ \\
Strain = $8\%$  & 3.44 & $\times$ & $\checkmark$ \\
\end{tabular}
\end{ruledtabular}
\end{table}




\section{Conclusion}

In summary, our first-principles investigations along with comprehensive symmetry analysis demonstrate that CrSb hosts a rich strain-induced topological landscape arising directly from its altermagnetic symmetry. By tuning either the electronic correlation strength $U_{\text{eff}}$ or applying isotropic tensile strain, we identify the emergence of symmetry-allowed Dirac and triple-point (TP) fermions along high-symmetry directions of the Brillouin zone. 
Further analysis of the evolution of these quasiparticles under increased strain and $U_{\text{eff}}$ reveals that the Dirac crossings are more sensitive to the tuning parameters. As the isotropic tensile strain is increased beyond a critical regime, the Dirac point annihilates while the TP fermion pair remains robust, indicating that strain selectively stabilizes the triple-point crossing. This behavior highlights that, although moderate electron localization is essential for the appearance of both quasiparticle types, the triple-point fermions are ultimately protected by the underlying $C_{3v}$ symmetry of the strain-preserved little group, while the Dirac state is not. Having previously identified Weyl fermions intrinsic to the altermagnetic phase of CrSb~\cite{Li2025,Lu2025}, our present results further reveal that distinct classes of nontrivial quasiparticles can be stabilized or suppressed through symmetry-preserving external control. These findings therefore establish isotropic tensile strain as a reliable and effective topological selector in altermagnetic material. 
Furthermore, the isotropic tensile strain proposed in this work can be experimentally realized through the application of negative chemical pressure, for example by partially substituting either the magnetic Cr sites or the nonmagnetic Sb sites with slightly larger atoms. Such targeted alloying-induced expansion provides a feasible pathway for tuning the lattice in real materials \cite{Singh2018}, and we leave the detailed exploration of these experimental strategies to future work. Overall, our findings expand the scope of altermagnetism by showing that strain engineering offers a powerful and previously underexplored route to stabilizing nontrivial topological quasiparticles in altermagnetic materials. This work, therefore, opens the door to a broader search for topological phases in altermagnets, guided by controllable structural perturbations such as strain.
\begin{acknowledgments}
The use of a high-performance computing facility at IISER Bhopal is gratefully acknowledged.
\end{acknowledgments}
%
\end{document}